\documentclass[prl,twocolumn,showpacs,floats]{revtex4}

\usepackage{graphicx}

\begin{document}

\title{Superconducting Gap and Pseudogap in Sm(O$_{1-x}$F$_x$)FeAs Layered Superconductor
from Photoemission Spectroscopy}
\author{Haiyun Liu$^{1}$, Xiaowen Jia$^{1}$, Wentao Zhang$^{1}$, Lin Zhao$^{1}$, Jianqiao Meng$^{1}$,
Guodong  Liu$^{1}$, Xiaoli Dong$^{1}$, G. Wu$^{2}$, R. H. Liu$^{2}$,
X. H. Chen$^{2}$, Z. A. Ren$^{1}$, Wei Yi$^{1}$, G. C. Che$^{1}$, G.
F. Chen$^{3}$, N. L. Wang$^{3}$,  Guiling Wang$^{4}$, Yong
Zhou$^{4}$, Yong Zhu$^{5}$, Xiaoyang Wang$^{5}$, Zhongxian
Zhao$^{1}$, Zuyan Xu$^{4}$, Chuangtian Chen$^{5}$, X. J.
Zhou$^{1,*}$}

\affiliation{
\\$^{1}$National Laboratory for Superconductivity, Beijing National Laboratory for Condensed
Matter Physics, Institute of Physics, Chinese Academy of Sciences,
Beijing 100190, China
\\$^{2}$Hefei National Laboratory for Physical Sciences at Microscale and Department of Physics,
University of Science and Technology of China, Hefei, Anhui 230026,
China
\\$^{3}$Beijing National Laboratory for Condensed Matter Physics, Institute of Physics,
Chinese Academy of Sciences, Beijing 100190, China
\\$^{4}$Key Laboratory for Optics, Beijing National Laboratory for Condensed Matter Physics,
Institute of Physics, Chinese Academy of Sciences, Beijing 100190,
China
\\$^{5}$Technical Institute of Physics and Chemistry, Chinese Academy of Sciences, Beijing 100190, China
}
\date{May 25, 2008}
%
%

\begin{abstract}

High resolution photoemission measurements have been carried out on
non-superconducting  SmOFeAs parent compound and superconducting
Sm(O$_{1-x}$F$_x$)FeAs (x=0.12, and 0.15) compounds. The
momentum-integrated spectra exhibit a clear Fermi cutoff that shows
little leading-edge shift in the superconducting state which
suggests the Fermi surface sheet(s) around the $\Gamma$ point may
not be gapped in this multiband superconductors.  A robust feature
at 13 meV is identified in all these samples. Spectral weight
suppression near E$_F$ with decreasing temperature is observed in
both undoped and doped samples that points to a possible existence
of a pseudogap in these Fe-based compounds.

\end{abstract}


\maketitle

The recent discovery of superconductivity in iron-based
oxypnictides\cite{Kamihara,GFChenLa, HHWen,
XHChen43K,GFChenCe,ZARenNd,ZARenPr,ZARenSm,XHPhaseDiagram}  has
attracted much attention because of their unusual high
superconducting transition temperature (T$_c$). The highest T$_c$
achieved so far (55 K)\cite{ZARenSm} has exceeded the
generally-believed limit ($\sim$40 K) set by the traditional BCS
theory of superconductivity, putting them into the second class of
``high temperature superconductors" in addition to the
cuprates\cite{Bednorz}. The parent compound of these iron-based
superconductors has been found to exhibit a spin density
wave(SDW)-like transition near 150 K and a possible
antiferromagnetic ground
state\cite{DongSDW,XHPhaseDiagram,PCDai,McGuireNeutron}. Doping
charge carriers into the system induces superconductivity at
appropriate doping levels. These behaviors appear to be similar to
those in cuprate superconductors.  One natural question to ask is
whether the superconductivity mechanism in these new Fe-based
superconductors is unconventional, and whether it may share the same
mechanism as that in the cuprate superconductors.

There are a couple of related issues that need to be addressed about
these new Fe-based superconductors. One is about the ground state of
undoped parent compound and how strong the electron correlation is.
The second is about the superconducting gap symmetry and the pairing
mechanism. The third is on whether its normal state is anomalous and
whether there is a pseudogap in the normal state as found in the
cuprate superconductors\cite{PseudogapReview}. Photoemission
spectroscopy, as a powerful tool to  directly measure the electronic
structure and energy gap, can shed important light on these
issues\cite{ARPESReview}.

\begin{figure}[tbp]
\begin{center}
\includegraphics[width=0.85\columnwidth,angle=0]{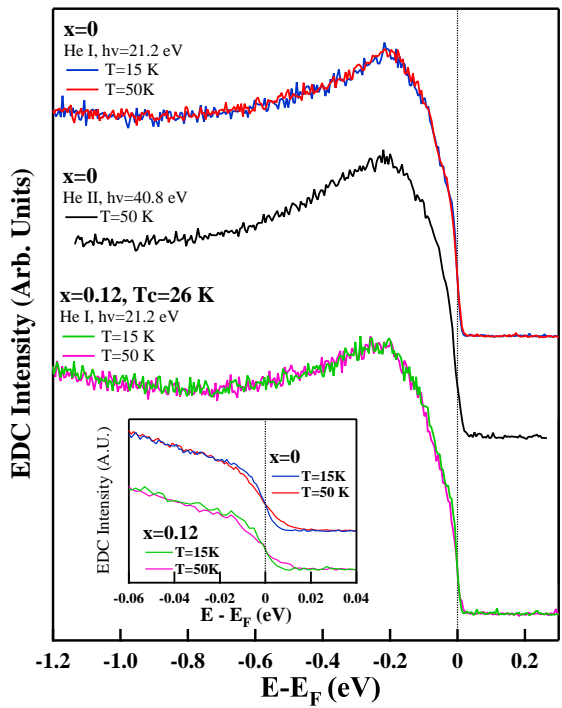}
\end{center}
\caption{Photoemission spectra of the Sm(O$_{1-x}$F$_x$)FeAs (x=0
and 0.12) samples measured at different temperatures using 21.2 eV
and 40.8 eV photon energies from the Helium lamp . The inset shows
the near-E$_F$ region for the samples measured at 15 K and 50 K
using 21.2 eV photon energy. }
\end{figure}

In this paper, we report high resolution photoemission measurements
on the Sm(O$_{1-x}$F$_x$)FeAs compounds at various doping levels. We
have found that the momentum-integrated spectra exhibit a clear
Fermi cutoff with little leading-edge shift in the superconducting
state. This suggests that, in this multiband superconductor,
different Fermi surface sheets may have different superconducting
gaps and the Fermi surface sheet(s) around the $\Gamma$ point may
not be gapped. We have also identified a robust feature at 13 meV
that is present in all these samples with different doping levels.
Near-E$_F$ spectral weight suppression with decreasing temperature
is observed in both undoped and doped samples. It points to a
possible existence of a pseudogap in these Fe-based compounds.

The photoemission measurements have been carried out on our
newly-developed system using both Vacuum Ultraviolet (VUV) laser and
 Helium discharge lamp as light sources\cite{GDLiu}. The advantages
of the VUV laser photoemission lie in its super-high energy
resolution, bulk sensitivity and high photon flux to give high
statistics of the data\cite{GDLiu}. The photon energy of the VUV
laser is 6.994 eV with a bandwidth of 0.26 meV. For the laser
measurements, the energy resolution of the electron energy analyzer
(Scienta R4000) was set at 1.5 meV, giving an overall energy
resolution of 1.52 meV. The spot size of the laser is less than 0.2
mm. The Helium lamp can provide two photon energies at 21.218 eV
(Helium I resonance line) and 40.813 eV (Helium II resonance line).
The energy resolution for the 21.218 eV measurements was set at 7.5
meV while for the 40.813 eV measurements at 20$\sim$30 meV.

\begin{figure}[tbp]
\begin{center}
\includegraphics[width=0.85\columnwidth,angle=0]{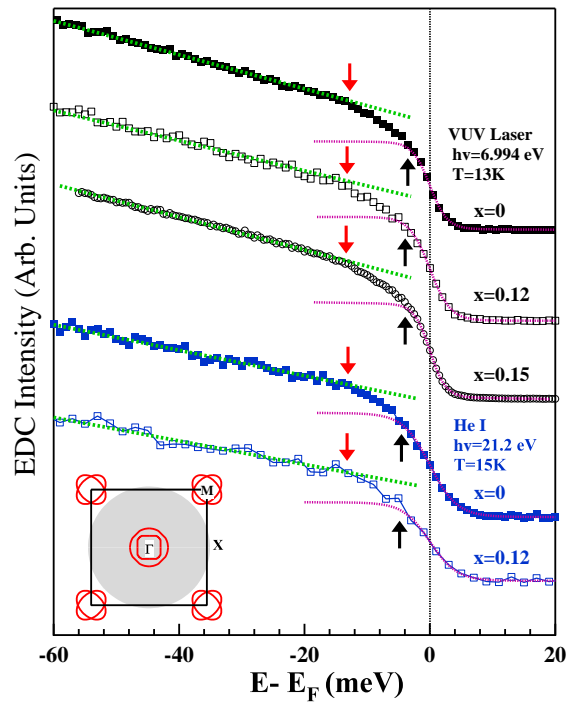}
\end{center}
\caption{Photoemission spectra of Sm(O$_{1-x}$F$_x$)FeAs (x=0, 0.12,
and 0.15) samples taken with laser (6.994 eV) and Helium lamp (21.2
eV) at 13$\sim$15 K. The high binding energy part between 20 and 60
meV shows linear behavior, as indicated by the fitted dashed lines.
The curves start to deviate from the linear line at $\sim$13 meV, as
indicated by the red down arrows. Upon approaching the Fermi level,
another drop occurs at $\sim$4 meV, as marked by the black up
arrows. The spectra near the Fermi level can be well fitted using
Fermi-Dirac distribution function, as indicated by the purple dotted
curves. The inset shows a schematic Brillouin zone and the
calculated Fermi surface\cite{DongSDW}. The momentum area that can
be covered by angle-integrated laser photoemission is marked as a
shaded region. }
\end{figure}

The polycrystalline Sm(O$_{1-x}$F$_x$)FeAs (x=0, 0.12 and 0.15,
nominal composition) samples are prepared by solid state reaction
method\cite{XHChen43K,XHPhaseDiagram,ZARenSm}. The x=0 sample is not
superconducting but with a possible SDW transition at $\sim$150 K,
while the x=0.12 and 0.15 samples are superconducting with T$_c$ at
26 K and 43 K, respectively\cite{XHPhaseDiagram}. For a
polycrystalline sample, in order to probe its intrinsic electronic
structure through photoemission measurements, great care has to be
taken to obtain clean surface and minimize the sample aging effect.
To get clean surface, we tried both scraping with a diamond file and
fracturing. We found that fracturing leads to cleaner surface, as
judged by the intensity and sharpness of the characteristic -0.2 eV
feature in the valence band (Fig. 1). Therefore, all samples in this
work were measured by fracturing {\it in situ} in vacuum with a base
pressure better than 5$\times$10$^{-11}$ Torr. In addition, we found
that these samples show aging effect even in the ultra-high vacuum,
manifested by depletion of the near-E$_F$ spectral weight and
particularly the change with time of the high binding energy part
in VUV laser photoemission spectra. Therefore, all the laser
photoemission data presented in the work were measured on a fresh
sample surface within 3$\sim$4 hours after sample fracturing.

\begin{figure}[tbp]
\begin{center}
\includegraphics[width=0.95\columnwidth,angle=0]{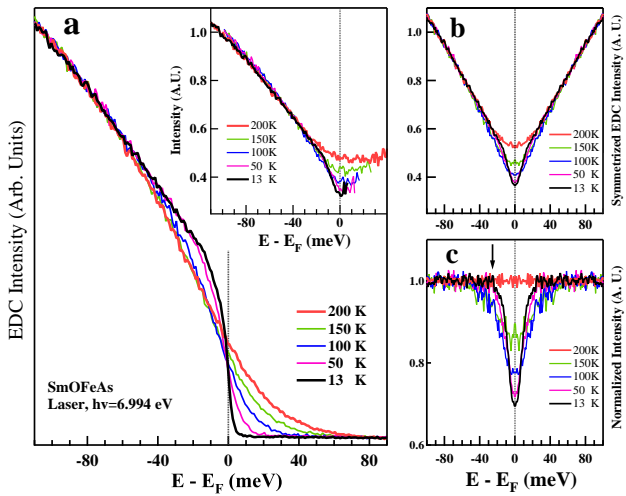}
\end{center}
\caption{(a). Temperature dependence of the laser photoemission
spectra on the SmOFeAs sample. The inset shows the same data but
with the Fermi-Dirac distribution function removed. (b). Symmetrized
spectra from (a) with respect to the Fermi level. (c). Symmetrized
spectra divided by the the data at 200 K. To avoid statistical
noise, the 200 K data is fitted with a polynomial and used for the
normalization. }
\end{figure}

Fig. 1 shows valence band of Sm(O$_{1-x}$F$_x$)FeAs with different
doping levels (x=0 and 0.12) and at different temperatures, measured
using different photon energies of the Helium lamp. The main feature
in these compounds is the peak near -0.2 eV\cite{HWOu}. Different
photon energy gives similar spectra, as seen from the 21.2 eV and
40.8 eV measurements on the x=0 sample. There is little spectral
change with temperature except for the near-E$_F$ region, as shown
in the inset of Fig. 1. There is no obvious valence band change
observed between different samples (x=0 and 0.12), specifically, the
-0.2 eV peak shows little position shift with doping.

Fig. 2 shows photoemission spectra of Sm(O$_{1-x}$F$_x$)FeAs (x=0,
0.12, and 0.15) measured using VUV laser at 13 K, together with data
taken at 15 K using 21.218 eV photon energy from the Helium lamp.
Two obvious kink features can be identified from these data. One is
at $\sim$13 meV where the spectrum starts to deviate from the linear
behavior at high binding energy. The other is at lower binding
energy near 4 meV that is due to the Fermi function cutoff. The
super-high energy resolution and low temperature make the 13 meV
feature well separated from the Fermi cutoff and clearly visible. We
note that this 13 meV feature is robust because it is present in
both undoped sample and doped samples, and seen in both laser
photoemission data and high resolution helium lamp measured data.

The high resolution and low temperature data in Fig. 2 make it
possible to examine possible superconducting gap in the x=0.12 and
x=0.15 superconductors. Generally speaking, in a momentum-integrated
photoemission spectrum, an s-wave superconducting gap on a single
Fermi surface can be easily identified from the leading-edge shift,
as demonstrated in Boron-doped superconducting diamond where the
leading-edge shift is visible even for a superconducting gap less
than 1 meV\cite{DiamondGap}. But for a system with multiple
superconducting gaps on different Fermi surface sheets, the
situation becomes not so straightforward. The leading edge position
in this case is mainly dictated by the minimum superconducting gap
on a Fermi surface sheet. The larger energy gaps on the other Fermi
surface sheets will then show up as additional features in the
spectrum at higher binding energy. For a superconductor with two
s-wave gaps like MgB$_2$\cite{MgB2Gap}, both the leading edge shift
and high binding energy feature are resolved to represent the
multiple gap structures. The situation will get more complicated if
non-s-wave gaps are involved on different Fermi surface sheets
and/or superconducting coherence peaks are not well developed.

\begin{figure}[tbp]
\begin{center}
\includegraphics[width=0.95\columnwidth,angle=0]{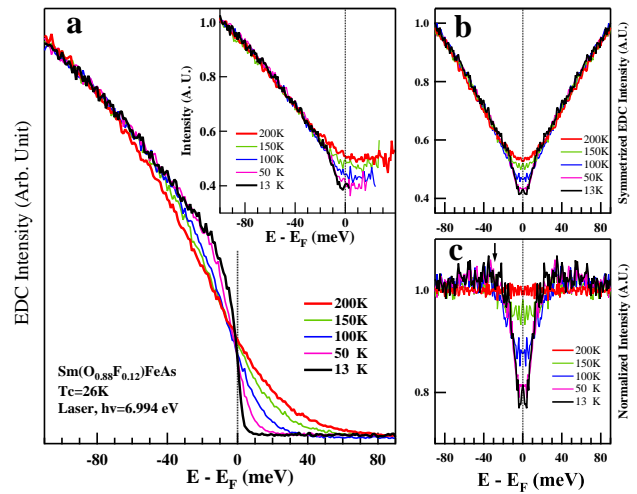}
\end{center}
\caption{(a). Temperature dependence of the laser photoemission
spectra on the Sm(O$_{1-x}$F$_x$)FeAs (x=0.12, T$_c$=26 K) sample.
The inset shows the same data but with the Fermi-Dirac distribution
function removed. (b). Symmetrized spectra from (a) with respect to
the Fermi level. (c). Symmetrized spectra divided by the data 200 K.
To avoid statistical noise, the 200 K data is fitted with a
polynomial and used for the normalization. }
\end{figure}

The Fe-based compounds have multiple Fermi surface sheets around
$\Gamma$ point and M($\pi$, $\pi$) point, as shown in the inset of
Fig. 2\cite{DongSDW}. It is possible that these Fermi surface sheets
may have different superconducting order parameters. As seen in Fig.
2, there is no obvious coherence peaks developed on the spectra in
the superconducting state, even for the x=0.15 sample with a T$_c$
as high as 43 K. This lack of coherence may come from disorder or
unconventional pairing symmetry. Particularly, the leading edge of
the spectra shows minimal shift from the Fermi level, as seen from
the data in Fig. 2 for both x=0.12 sample with T$_c$=26 K and x=0.15
sample with T$_c$=43 K. In fact, the Fermi cutoff for these samples
can be well fitted with a Fermi-Dirac distribution function
convoluted with the energy resolution used (Fig. 2), indicating a
nearly zero leading-edge shift. Supposing this zero-shift Fermi edge
is not from non-superconducting metallic impurities in the samples,
it indicates that there are ungapped Fermi surface sheet(s) in the
Sm(O$_{1-x}$F$_x$)FeAs superconductors. We note that with a laser
photon energy at 6.994 eV,  the momentum space it can cover does not
span the entire Brillouin zone: it covers the Fermi surface sheets
around $\Gamma$ point, but not those around the M($\pi$,$\pi$) point
(inset of Fig. 2).  The laser data then further suggest that there
are ungapped Fermi surface sheet(s) near the $\Gamma$ point.

It is tempting to see whether the $\sim$13 meV feature might
represent a superconducting gap in the Sm(O$_{1-x}$F$_x$)FeAs
superconductors. However, this possibility can be simply ruled out
because it is also observed  in the undoped x=0 sample that is not
superconducting. We do not see other obvious features from the laser
data between the Fermi level and 13 meV (Fig. 2) that can be taken
as signatures of the superconducting gap on other Fermi surface
sheets. To probe the superconducting gap on the Fermi surface sheets
near the M($\pi$,$\pi$) point (inset of Fig. 2), one can in
principle rely on high photon energy data that can cover the entire
Brillouin zone, like the Helium lamp data in the inset of Fig. 1.
However, given that the Fermi surface sheet(s) around the $\Gamma$
point is not gapped in the superconducting state as suggested from
our laser data, it will also give rise to a zero-shifted leading
edge in the high photon energy data. This would make it difficult to
extract the gap information from the leading edge in the high photon
energy spectrum. Again, one has to look for signatures at higher
binding energy in the Helium lamp (or other high photon energy)
spectra that require high energy resolution and high data
statistics.

Fig. 3 shows detailed temperature dependence of the laser
photoemission spectra for the undoped SmOFeAs sample.  The
temperature-induced change is mainly confined near the Fermi level
region; the high binding energy spectra at different temperatures
can be normalized to overlap with each other. Note that the spectra
at different temperatures do not cross the same point at the Fermi
level, a behavior that is different from a normal metal like gold
where all spectra cross at the same energy E$_F$. To remove the
effect of Fermi cutoff, the spectra are divided by Fermi-Dirac
distribution function at the respective temperature and shown in the
inset of Fig. 3. One can see a suppression of the spectral weight
near the Fermi level with decreasing temperature, a behavior that
starts at high temperatures even at 150 K. This behavior is similar
to the normal state spectral weight depletion near the antinodal
region in the underdoped cuprate superconductors which is related to
the opening of a pseudogap\cite{PseudogapReview}.

To further examine the possible opening of the pseudogap in Fe-based
compounds, we follow the procedure that is commonly used in
high-T$_c$ cuprate superconductors\cite{NormanSymmetrize} by
symmetrizing the original data in Fig. 3a with respect to the Fermi
level (Fig. 3b). This is another way to remove the Fermi-Dirac
distribution function and it provides a visualized way to look for a
gap. Again one sees clearly the depletion of spectral weight near
the Fermi level with decreasing temperature (Fig. 3b).  To highlight
the effect caused by temperature, we further divide the symmetrized
spectra with the one at 200 K (Fig. 3c).  The suppression of the
spectral weight near the Fermi level becomes clearer and one can now
identify an energy scale at which the spectral weight starts to
lose. It is $\sim$25 meV for the 13 K data, shows slight increase
with increasing temperature, but overall lies in the 25$\sim$40 meV
energy range for different temperatures (Fig. 3c).

The temperature dependence of spectra for the x=0.12 superconducting
sample (T$_c$=26 K) (Fig. 4) appears to be surprisingly similar to
that in the undoped SmOFeAs (Fig. 3). Here again one sees
suppression of spectral weight with decreasing temperature as in the
inset of Fig. 4a, symmetrized data in Fig. 4b and normalized data in
Fig. 4c.  For this superconducting sample with T$_c$=26 K, one may
wonder whether the near-E$_F$ spectral weight suppression in the
superconducting state, like the 13 K  data in Fig. 4, compared with
a normal state data at 50 K, can be taken as a signature of
superconducting gap opening.  Note that the same behavior also
occurs in the non-superconducting SmOFeAs sample (Fig. 3), we
believe this is not a reliable way in judging on a superconducting
gap.

In summary, from our high resolution photoemission measurements on
the Sm(O$_{1-x}$F$_x$)FeAs compounds, we found zero leading-edge
shift in the spectra in the superconducting state. This suggests
that the Fermi surface sheet(s) around the $\Gamma$ point may not be
gapped. We have identified a robust feature at 13 meV in different
samples. Whether this could be caused by electron coupling with some
bosonic mode needs to be further investigated.  Spectral suppression
near E$_F$ with decreasing temperature is observed in both undoped
and doped samples. This points to possible existence of a pseudogap
in the Fe-based compounds. Whether this is caused by local SDW
fluctuation or strong electron-boson coupling needs further
experimental and theoretical studies.

This work is supported by the NSFC, the MOST of China (973 project
No: 2006CB601002, 2006CB921302), and CAS (Projects ITSNEM).

$^{*}$Corresponding author: XJZhou@aphy.iphy.ac.cn


\end{document}